\documentclass[12pt,dvips]{article}
\usepackage{epic}
\usepackage{eepic}
\usepackage{graphics}
\usepackage{amsfonts}
  \usepackage{amsthm}
 \usepackage{amsmath}
 \usepackage{amscd}
 \usepackage{amssymb}
 \usepackage{latexsym}

\numberwithin{equation}{section}
 \newtheorem{thm}{Theorem}[section]
 \newtheorem{lm}{Lemma}[section]
 \newtheorem{prop}{Proposition}[section]
\theoremstyle{definition}

 \title{\bf Non-uniform continuity of periodic Holm-Staley b-family of equations}
\author{Ognyan Christov$^1$, Sevdzhan Hakkaev$^2$, Iliya D. Iliev$^3$}
\date{}
\begin{document}

\maketitle

\noindent
{\small
$^1$ Faculty of Mathematics and Informatics, Sofia University, 1164 Sofia, Bulgaria\\
$^2$ Faculty of Mathematics and Informatics, Shumen University,
      9712 Shumen, Bulgaria \\
$^3$ Institute of Mathematics, Bulgarian Academy of Sciences,  Bl.
8, 1113 Sofia, Bulgaria  }

\vspace{2ex}
\noindent
\begin{abstract}
\noindent
We consider a family of non-evolutionary partial differential
equations known as Holm - Staley b - family which includes the
integrable Camassa-Holm and Degasperis-Procesi equations. We show
that the solution map is not uniformly continuous. The proof
relies on a construction of smooth periodic travelling waves with
small amplitude.
\end{abstract}

\section{Introduction}

In \cite{HS1, HS2} D. Holm and M. Staley studied an one-dimensional
version of an active fluid transport that is described by the
following nonlinear equation
\begin{equation}
 \label{0.1}
    m_{t} + u m_{x} + b u_{x} m = 0,
\end{equation}
with $m = u - u_{xx}$, $u (x, t)$ representing the fluid velocity,
while the constant $b$ is a balance or a bifurcation parameter for the
solution behavior.  It has been shown in \cite{DP} that
equation (\ref{0.1}) is integrable only for $b = 2$ and $b = 3$.

In this paper we study the periodic Cauchy problem for the b -
family of equations (\ref{0.1}), namely
\begin{equation}
 \label{0.4}
    u_{t} - u_{xxt}  + (b + 1) u u_{x} = b u_x u_{xx} + u u_{xxx},
    \quad
    u (0) = u_0, \quad t \geq 0, \, x \in \mathbb{S}.
\end{equation}
If $b = 2$, then (\ref{0.4}) becomes the Camassa-Holm (CH) equation
 \begin{equation}
 \label{0.2}
   u_{t} - u_{xxt} + 3u u_{x} = 2 u_{x} u_{xx} + u u_{xxx}.
 \end{equation}
 Our aim here is to enlarge the result of Himonas and Misiolek \cite{HM}
(originally proved for the CH equation) for all real $b \neq 0$.

Equation (\ref{0.2}) was first derived by Fokas and Fuchssteiner \cite{FF} as a
bi-Hamiltonian system, and then by Camassa and Holm \cite{CaHo} as a model
for shallow water waves. The Cauchy problem for the CH equation in both
periodic and non periodic case was studied extensively. It has been shown that
the Camassa-Holm equation is locally well-posed in $ \mathrm{H}^s,\;s >
\frac{3}{2}$ with solutions depending continuously on initial data
\cite{Co1,CoEs2, CoEs5,LiOl, Ro}. The Camassa-Holm equation has
global solutions but also solutions which blow-up in finite time
(see \cite{Co1, Co3, Co4, CoEs1, CoEs2,CoEs5, Zh}).

 When $b = 3$ in (\ref{0.4}), we recover Degasperis-Procesi
 (DP) equation which is a model for nonlinear shallow water dynamics,
 \begin{equation}
   \label{0.3}
   u_{t} - u_{xxt} + 4 u u_{x} = 3 u_{x} u_{xx} + u u_{xxx}.
 \end{equation}
 The Cauchy problem for the DP equation in both periodic and
 non periodic case is studied in \cite{ELY, LY,Yin4, Yin5, Zh2}. For the
 equations (\ref{0.2}) and (\ref{0.3}) the blow-up occurs as wave
 breaking, that is, the solutions remains bounded but its slope
 becomes infinite in finite time.

 Sometimes, it is more appropriate to consider other version of
 well-posedness problem, for example if one strengthens the notion
 of well-posedness, requiring that the mapping data-solution is
 uniformly continuous. The ill-posedness of some classical  nonlinear
 dispersive equations (for instance Korteweg-de Vries  equation,
 modified Korteweg-de Vries equation, cubic Schr\"odinger  equation,
 and Benjamin-Ono equation) in both periodic and  non-periodic cases
 are studied in  \cite{AH, BKPS, BPS}. The approach in  these papers is
 based on the existence and suitable properties of the traveling wave solutions
 associated to the equations. In particular,  a good behavior of their Fourier
 transforms is required. In \cite{HM}  Himonas and Misiolek showed that for
 $s \geq 2$ the solution map $u_{0}  \rightarrow u$ for the CH equation is
 not uniformly continuous from any  bounded set in $\mathrm{H}^{s}(\mathbb{S})$ into
 $C([0,T] , \mathrm{H}^{s}(\mathbb{S}))$.  A key step in the proof of that result is a
 construction of a sequence of  smooth travelling waves. Himonas,
 Kenig and Misiolek \cite{HKM} extend the result to the range
 $\frac{3}{2} < s < 2$. Their proof is based on the approximation of
 solutions by terms containing  high and low frequencies  and
 exploring the conservation of the $\mathrm{H}^1$ norm. Note that
 $\mathrm{H}^1$ norm is  a conservation law for equation
 (\ref{0.4}) only for $b = 2$.

 Recently Gui, Liu, and Tian \cite{GLT} considered the equation  (\ref{0.1})
 on the real line. They proved that the equation is  locally well-posed in
 the Sobolev space $\mathrm{H}^{s}(\mathbb{R})$ for  $s > {\frac{3}{2}}$. Moreover,
 they gave the precise blow-up  scenario of strong solution of the equation
 with certain initial  data. In \cite{Zh1} Zhou established blow-up results
 for this family of equations  under  various classes of initial data. He
 also proved that the solutions with compact support initial data do not
 have compact support. In the periodic case, sufficient conditions on the
 initial data are obtained in \cite{ChHa} to guarantee the finite time
 blow-up and global existence. Using the ideas from \cite{HM}, it is also
established there the non-uniform continuity of DP equation.

Our main result in this paper is the following.
\begin{thm}
\label{thm1}
For any $s \geq 3$, the solution map $u_0 \to u$ for
the  equation (\ref{0.4}) with $b \neq 0$, is not uniformly continuous
from any
bounded set in $\mathrm{H}^s (\mathbb{S})$ into $\mathrm{C}([0,
t_0], \mathrm{H}^s (\mathbb{S}))$. More precisely, for each $s \geq
3$ there exist constants $c_{1, 2} > 0$ and two sequences of smooth
solutions $u_n, v_n$ of the equation (\ref{0.4}) such that for any
$t \in [0, 1]$
$$
\sup_n ||\, u_n (t) ||_{\mathrm{H}^s} + \sup_n || v_n (t)
||_{\mathrm{H}^s} \leq c_1,
$$
$$
\lim_{n \to \infty}  ||\, u_n (0) - v_n (0)
||_{\mathrm{H}^s} = 0,
$$
$$
\liminf_n ||\, u_n (t) - v_n (t) ||_{\mathrm{H}^s} \geq c_2 \sin
\left(\frac{t}{2}\right).
$$
\end{thm}

\vspace{2ex}

The idea of the proof is borrowed from \cite{HM}.
Two sequences of exact periodic smooth solutions are constructed  taking advantage
of a scaling property of the b - family. While their initial states converge
in $\mathrm{H}^s$ - norms, the
solutions remain apart at certain time.  We use two different parameters
but equivalent to those in \cite{HM}
 in order to define appropriate families of solutions.
The careful choice of these two parameters is crucial in deriving the
$\mathrm{H}^s$ estimates.
Due to the transcendent dependence on $b$, here we do not give the sharp estimates for
these parameters and merely say that they are sufficiently small.

The paper is organized as follows. In section 2 the periodic
travelling waves of (\ref{0.4}) are studied. Although the
corresponding conservative system describing the travelling waves
is somehow transcendent and depends on several parameters, the
things are arranged so that we study an equivalent Hamiltonian
quadratic system for which the conditions for the existence of
periodic solutions are more or less known.

The main difficulty here is to establish estimates for the period.
This is done in section 3 by calculating the first two terms in
the expansion of the period function for periodic travelling waves
with small amplitude.

In section 4 we obtain upper estimates for these solutions in $\mathrm{H}^s$ -
 norm and carry on the proof of Theorem \ref{thm1} for $ b \neq 0, \pm 1$.

We summarize the corresponding results for the case $b = \pm 1$ in section 5.
This approach is not applicable to the case $b = 0$
due to lack of periodic solutions.

\section{Periodic travelling waves}

In this section we investigate the periodic travelling wave
solutions of the b - family equation
\begin{equation}
\label{1.1}
 u_{t} - u_{xxt}  + (b + 1) u u_{x} = b u_x u_{xx} + u u_{xxx}  ,\quad b \in \mathbb{R}.
\end{equation}
Note that if $u (x, t)$ is a classical solution of (\ref{1.1}), then such is
the function
$$
u_c (x, t) = c u(x, ct), \quad \mbox{for any constant $c$}.
$$
First, take velocity $c = 1$ and look for a travelling-wave
solution of (\ref{1.1}) of the form  $u (x, t) = \varphi(x - t)$.
One can integrate twice the respective equation
\begin{equation}
\label{fi}
-\varphi'+\varphi'''+(b+1)\varphi \varphi'=b\varphi'\varphi''+\varphi\varphi'''
\end{equation}
to obtain, if $b\neq 1$,
\begin{equation}
\label{1.2}
|1 - \varphi|^{b - 1}\left[\varphi'^2 - \varphi^2
+ \frac{2C_1}{b - 1} \right] = 2C_2,
\end{equation}
and, if $b=1$,
\begin{equation}
\label{b=1}
\varphi'^2 - \varphi^2
+ 2C_1\ln|1-\varphi| = 2C_2,
\end{equation}
where $C_1,C_2$ are constants of integration. Take the general case $b\neq 1$.

In the $(X, Y)$-plane with $X = \varphi$, $Y = \varphi'$ consider the
autonomous system
\begin{equation}
\label{1.3}
\begin{array}{l}
\dot{X} = H_Y/M = 2Y(1 - X),\\
\dot{Y} = -H_X/M = 2X(1 - X) + (b - 1)(Y^2 - X^2 + d),
\end{array}
\end{equation}
having a first integral $H$ and an integrating factor $M$, as follows:
\begin{equation}
  \label{1.4}
   H(X, Y) = |\,1-X|^{b-1}(Y^2 - X^2 + d),\quad M (X) = (1-X)|\, 1 - X|^{b-3},
\end{equation}
respectively, where it is taken for short $d = 2 C_1(b-1)^{-1}$.
As well known, system (\ref{1.3}) has a periodic solution if and only if
it has a center. The coordinates $(X,Y)$ of a center of (\ref{1.3}) must
satisfy
\begin{equation}
\label{root}
(1 + b) X^2 - 2 X + (1 - b)d = 0,\;\; Y = 0; \;\; [1-X][1-(b+1)X]<0.
\end{equation}
One can easily verify the following statement.

\vspace{2ex}

\noindent
\begin{prop}
\label{p11}
 Let $b\neq 0,\pm 1$. System (\ref{1.3}) has a center if and only if one of the
following conditions holds:

\vspace{1ex}
(i) $\quad|\, b| > 1,\quad$ $\displaystyle \frac{1}{1-b^2} < d < 1.$

\vspace{1ex}
(ii) $\quad|\, b| < 1,\quad$ $\displaystyle 1 < d < \frac{1}{1-b^2}.$

\vspace{1ex}
(iii) $\quad b < -1,\quad$ $d \geq 1.$

\end{prop}
\vspace{1ex}
\noindent
We observe that  $\Delta = 1 + d (b^2 - 1) > 0$ for all cases.
See the corresponding phase portraits of the systems with a center on Figure 1.
Note that, by (\ref{root}), there are no periodic orbits in (\ref{1.3})
if $b=0$. Besides, cases $b=\pm1$ will be considered separately in Section 5.
Therefore, we will assume below that $b \neq 0, \pm1$.
\begin{figure}
\label{fig1}
\begin{center}
 \setlength{\unitlength}{1.9mm}

 \begin{picture}(0,45)(0,0)
 \linethickness{0.4pt}

       \qbezier(-9,40)(0,0)(9,40)  
       \qbezier(21,-15)(27,-5)(35,-3)  
       \qbezier(-21,-15)(-27,-5)(-35,-3)  
       \qbezier(-35,20)(15,20)(34,20)  

       \put(1,19){\makebox(0,0)[cc]{${\scriptstyle 1}$}}
       \put(-17,3){\makebox(0,0)[cc]{${\scriptstyle -1}$}}
       \put(1,3){\makebox(0,0)[cc]{${\scriptstyle 0}$}}
       \put(17,3){\makebox(0,0)[cc]{${\scriptstyle 1}$}}

       \qbezier(0,43)(0.5,42)(1,41)  
       \qbezier(0,43)(-0.5,42)(-1,41)    
       \put(2,43){\makebox(0,0)[cc]{$d$}}

       \qbezier(-35,2)(15,2)(35,2)  
       \put(36,3){\makebox(0,0)[cc]{$b$}}
       \qbezier(35,2)(34,2.5)(33,3)   
       \qbezier(35,2)(34,1.5)(33,1)   
       \qbezier(0,-15)(0,38)(0,43)    
    \qbezier(18,-15)(18,38)(18,42)    
    \qbezier(-18,-15)(-18,38)(-18,42) 

 \linethickness{0.8pt}
\put(-31,42){\makebox(0,0)[cc] {${\scriptstyle I}$}}
\put(-13,42){\makebox(0,0)[cc] {${\scriptstyle II}$}}
\put(13,42){\makebox(0,0)[cc] {${\scriptstyle III}$}}

\put(31,18){\makebox(0,0)[cc] {${\scriptstyle V}$}}
\put(-31,18){\makebox(0,0)[cc] {${\scriptstyle IV}$}}

\put(31,2){\makebox(0,0)[cc] {${\scriptstyle VII}$}}
\put(-31,2){\makebox(0,0)[cc] {${\scriptstyle VI}$}}

\put(20,-11){\makebox(0,0)[cc] {${\scriptstyle IX}$}}
\put(-20,-11){\makebox(0,0)[cc] {${\scriptstyle VIII}$}}
 \end{picture}
$\ $
\end{center}

\begin{center}
 \setlength{\unitlength}{1mm}
 \begin{picture}(22,22)(-21,-64)
 \linethickness{0.4pt}
       \put(10,12){\circle*{1}}
       \put(16,12){\circle*{1}}
       \qbezier(0,2)(0,11)(0,20)

       \qbezier(15.4,12.7)(12.5,16)(10,16)

       \qbezier(10,16)(3,16)(3,12)
       \qbezier(10, 8)(3, 8)(3,12)
       \qbezier(15.4,11.3)(12.5, 8)(10, 8)

       \qbezier(16.4,12.7)(18,14)(20,15)

       \qbezier(16.4,11.2)(18, 9)(20,8.1)

 \linethickness{0.5pt}
    \qbezier(6.4,12)(6.4,14)(9.3,14)
    \qbezier(9.3,14)(12.5,14)(12.7,12)
    \qbezier(9.3,10)(12.5,10)(12.7,12)

    \qbezier(6.4,12)(6.4,10)(9.3,10)
 \end{picture}
 $\    $
 \setlength{\unitlength}{1mm}
 \begin{picture}(22,22)(-35,-64)
 \linethickness{0.4pt}
       \put(10,12){\circle*{1}}
       \put(4,12){\circle*{1}}
 \qbezier(20,2)(20,11)(20,20)

       \qbezier(4,12)(6.5,16)(9,16) 

       \qbezier(9,16)(16,16)(16,12)
       \qbezier(9, 8)(16, 8)(16,12)
       \qbezier(4,12)(6.5,8)(9, 8) 

       \qbezier(4,12)(3,14)(2,15)

      \qbezier(4,12)(3,10)(2,9)

    \qbezier(6.4,12)(6.4,14)(9.3,14)
    \qbezier(9.3,14)(12.5,14)(12.7,12)
    \qbezier(9.3,10)(12.5,10)(12.7,12)

    \qbezier(6.4,12)(6.4,10)(9.3,10)

 \end{picture}
$\  $
 \setlength{\unitlength}{1mm}
 \begin{picture}(22,22)(70,-35)
 \linethickness{0.4pt}

 \linethickness{0.4pt}

      \put(6,11){\circle{7}}
       \put(6,11){\circle*{1}}
       \put(15,11){\circle*{1}}

       \put(20,17.1){\circle*{1}}
       \put(20,4.8){\circle*{1}}

       \qbezier(20,0)(20,11)(20,22)
       \qbezier(23,19)(20,18)(15,11)
       \qbezier(23,3.5)(20,4.5)(15,11)

       \qbezier(9,19)(12,18)(15,11)
       \qbezier(9,3.5)(12,4.5)(15,11)

       \qbezier(5,18)(17,10)(5, 4)
 \end{picture}
$\    $
 \setlength{\unitlength}{1mm}
\begin{picture}(22,22)(-5,-35)
 \linethickness{0.4pt}
       \put(14,11){\circle{5}}
       \put(14,11){\circle*{1}}
       \put(5,11){\circle*{1}}

       \put(20,18){\circle*{1}}
       \put(20,4.5){\circle*{1}}

       \qbezier(20,0)(20,11)(20,22)

       \qbezier(5,11)(8,15)(18,21)
       \qbezier(5,11)(4,10)(3,9)
       \qbezier(18,21)(18.5,21.5)(19,23)

       \qbezier(5,11)(8,7)(18,1)
       \qbezier(5,11)(4,12)(3,13)
        \qbezier(18,1)(18.5,0.5)(19,-1)

       \qbezier(9,11)(9.5,14)(23,19)
       \qbezier(9,11)(9.5,8)(23,3)
 \end{picture}

 \setlength{\unitlength}{1mm}
 \begin{picture}(22,22)(55,-90)
 \linethickness{0.4pt}
   \setlength{\unitlength}{1mm}
       \put(5,11){\circle{5}}
       \put(19,11){\circle{5}}
       \put(5,11){\circle*{1}}
       \put(19,11){\circle*{1}}

       \qbezier(12,2)(12,11)(12,20)

       \qbezier(19,19)(9.5,11)(19,3)
       \qbezier(3,19)(15.5,11)(3,3)

 \end{picture}

 \setlength{\unitlength}{1mm}
 \begin{picture}(22,22)(18,-60)
 \linethickness{0.4pt}
       \put(10,12){\circle*{1}}
       \put(16,12){\circle*{1}}
       \qbezier(18,4)(18,11)(18,18) 

       \put(18,9.2){\circle*{1}}
       \put(18,14){\circle*{1}}
       \qbezier(15.4,12.7)(12.5,16)(10,16)

       \qbezier(10,16)(3,16)(3,12)
       \qbezier(10, 8)(3, 8)(3,12)
       \qbezier(15.4,11.3)(12.5, 8)(10, 8)

       \qbezier(16.4,12.7)(18,14)(18,14)

       \qbezier(16.4,11.2)(18,9)(18,9)

      \qbezier(12,17)(18,11)(20,17)
      \qbezier(12,7)(18,11.3)(20,7)

 \linethickness{0.5pt}
    \qbezier(6.4,12)(6.4,14)(9.3,14)
    \qbezier(9.3,14)(12.5,14)(12.7,12)
    \qbezier(9.3,10)(12.5,10)(12.7,12)

    \qbezier(6.4,12)(6.4,10)(9.3,10)
 \end{picture}
 $\    $
 \setlength{\unitlength}{1mm}
 \begin{picture}(22,22)(-45,-60)
 \linethickness{0.4pt}
       \put(10,12){\circle*{1}}

       \put(4,12){\circle*{1}}
    \qbezier(17.5,2)(17.5,11)(17.5,20)

       \put(17.5,7){\circle*{1}}
       \put(17.5,17){\circle*{1}}

 \qbezier(17.5,7)(11.5,7)(4.5,5.5)  
 \qbezier(17.5,7)(19.5,7)(20,6)

 \qbezier(17.5,17)(11.5,17)(4.5,18.5)
 \qbezier(17.5,17)(19.5,17)(20,18.5)

       \qbezier(4,12)(6.5,16)(9,16) 

       \qbezier(9,16)(16,16)(16,12)
       \qbezier(9, 8)(16, 8)(16,12)
       \qbezier(4,12)(6.5,8)(9, 8) 

       \qbezier(4,12)(3,14)(2,15)

      \qbezier(4,12)(3,10)(2,9)

    \qbezier(6.4,12)(6.4,14)(9.3,14)
    \qbezier(9.3,14)(12.5,14)(12.7,12)
    \qbezier(9.3,10)(12.5,10)(12.7,12)

    \qbezier(6.4,12)(6.4,10)(9.3,10)

 \end{picture}
$\  $ \hspace{-26ex}
 \setlength{\unitlength}{1mm}
  \framebox{
 \begin{picture}(25,25)(0,0)
 \linethickness{0.4pt}
\put(1,1){\makebox(0,0)[cc] {${\scriptstyle VI}$}}

      \put(9,11){\circle{9}}
       \put(9,11){\circle*{1}}
       \put(15,11){\circle*{1}}
       \put(20,18){\circle*{1}}
       \put(20,4.5){\circle*{1}}

       \qbezier(20,0)(20,11)(20,22)

       \qbezier (7,0)(15,11)(23,22)
       \qbezier (7,22)(15,11)(23,0)

 \end{picture}
} $\    $
 \setlength{\unitlength}{1mm}
  \framebox{
 \begin{picture}(25,25)(0,0)
 \linethickness{0.4pt}
\put(1,1){\makebox(0,0)[cc] {${\scriptstyle VII}$}}
       \put(14,11){\circle{7}}
       \put(14,11){\circle*{1}}
       \put(5,11){\circle*{1}}
          \put(20,20){\circle*{1}}
       \put(20,2){\circle*{1}}

       \qbezier(20,0)(20,11)(20,22)

       \qbezier(-1,7)(5,11)(23,22)
       \qbezier(-1,14.66)(5,11)(23,0)

 \end{picture}
} $\    $

\end{center}

\caption{\hspace{-1.5pt} {\small Bifurcation diagram of system
(\ref{1.3}) in the $(b, d)$-plane. Phase portraits of the systems
having periodic solutions are shown only. The vertical invariant
line in the $(X, Y)$-space is always $X = 1$.}}
\end{figure}
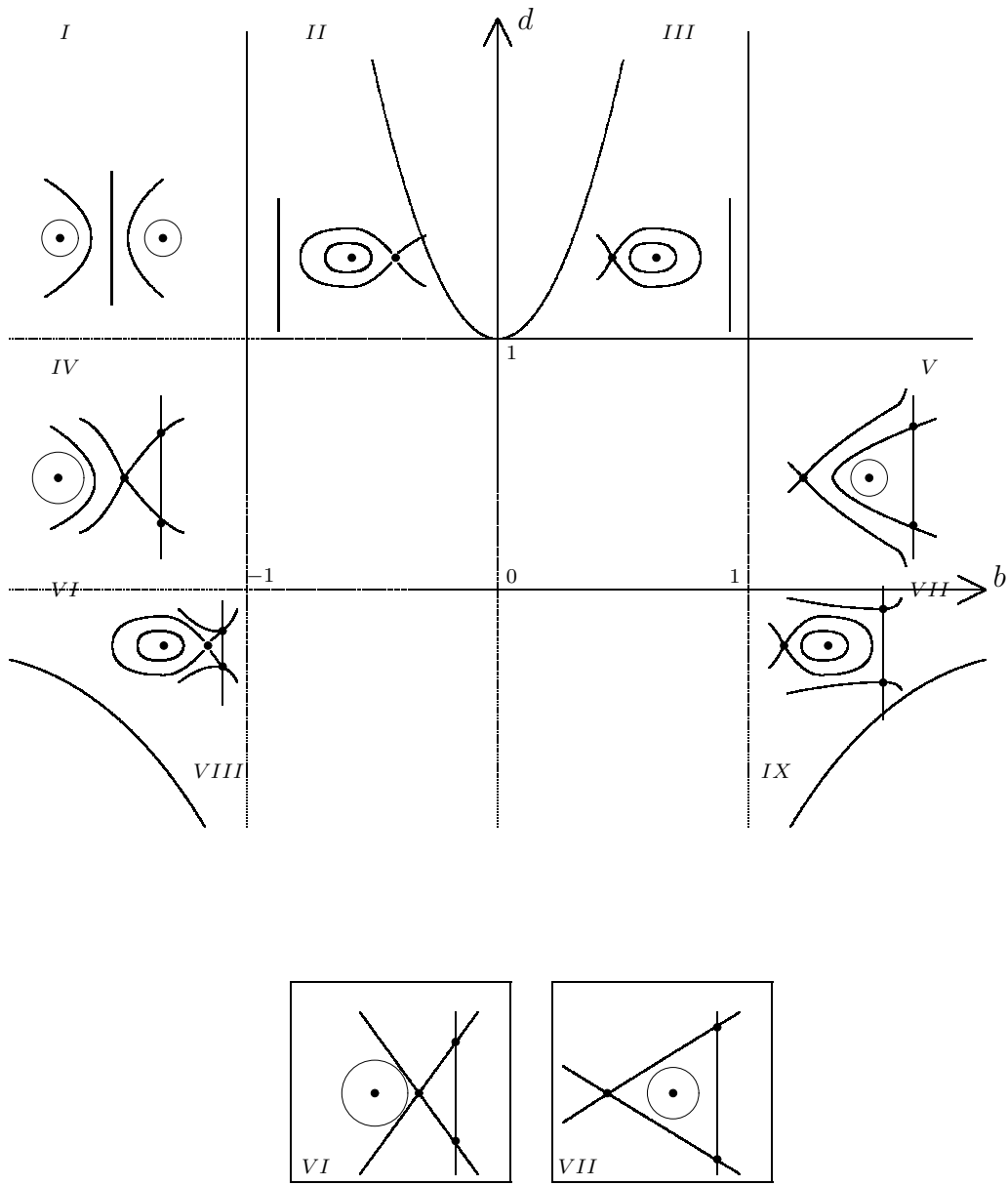

The types of quadratic centers are well known (see e.g. \cite{Z94}). Writing a
quadratic system with a center at the origin in the $(x, y)$-plane as
a complex equation with respect to $z = x + iy$, after rescaling we obtain the
following types \cite{BSM}:
$$
\begin{array}{lr}
\dot{z} = -iz - z^2  + 2|z|^2 + (A + iB)\bar{z}^2;& \mbox{\rm Hamiltonian,}\\
\dot{z} = -iz + Az^2 + 2|z|^2 + B\bar{z}^2;& \mbox{\rm Reversible,}\\
\dot{z} = -iz + 4z^2 + 2|z|^2 + (A + iB)\bar{z}^2,\;\;A^2 + B^2 = 4;&
\mbox{\rm Codimension 4,}\\
\dot{z} = -iz + z^2  + (A + iB)\bar{z}^2;& \mbox{Generalized Lotka-Volterra,}\\
\dot{z} = -iz + \bar{z}^2;& \mbox{Hamiltonian triangle}.
\end{array}
$$
In the equations above, $A$ and $B$ are real parameters.

By passing to the respective normal form, one
can prove the following structure result concerning the types of centers
in (\ref{1.3}).

\vspace{2ex}
\noindent
\begin{prop}
\label{p12}
Up to an affine transformation of the variables, the center of (\ref{1.3})
belongs to the following type:

\vspace{0.5ex}
(i) {\it Hamiltonian triangle, if $b = 2$, $d = 0$;}

\vspace{0.5ex}
(ii) {\it Lotka-Volterra, if $|\,b| > 1$, $b \neq 2$, $d = 0$, with
$\displaystyle (A, B) = \left(\frac{b}{b-2},0\right);$}

\vspace{0.5ex}
(iii) {\it Reversible, if  $d \neq 0$, $(b-\sqrt\Delta)(b+1)>0$, with}
$$(A,B)=\left(\frac{b\sqrt\Delta-4\sqrt\Delta+b}{b(\sqrt\Delta-1)},
\frac{\sqrt\Delta+1}{\sqrt\Delta-1}\right);$$

\vspace{0.5ex}
(iv) {\it Reversible, if  $d\neq 0$, $(b+\sqrt\Delta)(b+1)<0$, with}
$$(A,B)=\left(\frac{b\sqrt\Delta-4\sqrt\Delta-b}{b(\sqrt\Delta+1)},
\frac{\sqrt\Delta-1}{\sqrt\Delta+1}\right).$$
\end{prop}

We next proceed to determine the interval $\Sigma$ where the periodic
orbits exist. Namely, given $b$ and $d$ as in Proposition \ref{p11}, to find
the maximal open interval $\Sigma = \Sigma(b, d)$ such that
for any $e \in \Sigma$ the level curve
\begin{equation}
\label{1.5}
H(X, Y) = e
\end{equation}
contains an oval (a simple closed curve without critical points).
Clearly, one of the endpoints of $\Sigma$ is the level $e_c$ corresponding to
the center and the other is the level $e_s$ corresponding to the contour
at which the period annulus around the center terminates.

\vspace{1ex}
\noindent
\begin{prop}
  \label{p13}
The system (\ref{1.3}) has a periodic solution
for energy levels $e \in \Sigma = (e_c, e_s)$, where:
$$
\begin{array}{l}
\displaystyle \Sigma=\left(2\frac{|b-\sqrt\Delta|^{b-1}}{|b+1|^{b+1}}
(d(b+1)-\sqrt\Delta-1),0\right)\\[3mm]
\quad \mbox{\it for}\quad  b > 1,\; 0\leq d<1,\quad  \mbox{\it and for}\quad
b<-1,\;d\geq 0,\\[3mm]
\displaystyle \Sigma=\left(2\frac{|b-\sqrt\Delta|^{b-1}}{|b+1|^{b+1}}
(d(b+1)-\sqrt\Delta-1),
2\frac{|b+\sqrt\Delta|^{b-1}}{|b+1|^{b+1}}
(d(b+1)+\sqrt\Delta-1)\right)\\[3mm]
\displaystyle \quad  \mbox{\it for}\quad |b|>1,\; \frac{1}{1-b^2}<d<0
\quad  \mbox{\it and for}\quad 0<b<1, \;1<d<\frac{1}{1-b^2},
 \end{array}
$$
$$
\begin{array}{l}
 \displaystyle \Sigma=\left(2\frac{|b+\sqrt\Delta|^{b-1}}{|b+1|^{b+1}}
 (d(b+1)+\sqrt\Delta-1),
 2\frac{|b-\sqrt\Delta|^{b-1}}{|b+1|^{b+1}}
 (d(b+1)-\sqrt\Delta-1)\right)\\[3mm]
 \displaystyle \quad  \mbox{\it for}\quad -1<b<0,\; 1<d<\frac{1}{1-b^2},\\[3mm]
\displaystyle \Sigma=\left(2\frac{|b+\sqrt\Delta|^{b-1}}{|b+1|^{b+1}}
(d(b+1)+\sqrt\Delta-1),0\right)\\[3mm]
\quad \mbox{\it for}\quad  b < -1,\; d > 1.
\end{array}
$$
\end{prop}

And, finally, if $T = T (e)$, $e \in \Sigma$ is the (minimal)
period of the orbit contained in (\ref{1.5}), one can find the
limits $T_c(b, d) = \lim_{e \to e_c} T (e)$ and $T_s (b, d) =
\lim_{e \to e_s} T (e)$ ($T_s$ might be infinity). Then, for any
$T$ from the open interval with endpoints $T_c$ and $T_s$, there
will be (at least one) periodic orbit of (\ref{1.3}) having $T$ as
a period.

\vspace{2ex}
\noindent

\begin{prop}
\label{p14}
Let $x_c$ be the abscissa of a center of system (\ref{1.3}). Then
\begin{equation}
\label{Tc}
x_c = \frac{1\pm\sqrt\Delta}{1 + b},\qquad
T_c = 2\pi\sqrt\frac{1 - x_c}{(b + 1)x_c - 1}, \qquad T_s = \infty.
\end{equation}
\end{prop}


\section{The period function for small-amplitude travelling - wave solutions}

Below we calculate the first two terms in the expansion of the
period function in the case when the periodic wave $\varphi$ we study has
a small amplitude.

\newpage
\noindent
 \setlength{\unitlength}{1.1mm}
\begin{picture}(35,20)(-2,11)
 \linethickness{0.4pt}

       \qbezier(0,12)(15,12)(35,12)  

       \qbezier(35,12)(34,12.5)(33,13)   
       \qbezier(35,12)(34,11.5)(33,11)   
       \qbezier(0,5)(0,12)(0,19)    

       \qbezier(1,17)(0.5,18)(0,19)  
       \qbezier(-1,17)(-0.5,18)(0,19)    

 \linethickness{0.8pt}

\put(17,12){\ellipse{20}{10}}
\put(13,13.5){\makebox(0,0)[cc] {${\scriptstyle \varepsilon}$}}
\put(5.5,10.5){\makebox(0,0)[cc]  {${\scriptstyle x_0}$}}
\put(17,10.5){\makebox(0,0)[cc] {${\scriptstyle x_c}$}}
\put(15.5,11.7){\makebox(0,0)[cc] {${\cdot}$}}
\put(30,11.7){\makebox(0,0)[cc] {${\cdot}$}}
\put(30.5,10.5){\makebox(0,0)[cc]  {${\scriptstyle 1}$}}
\put(28,10.5){\makebox(0,0)[cc]  {${\scriptstyle x_1}$}}

\put(19.5,0.5){\makebox(0,0)[cc] {{Figure 2. The periodic solution}}}
 \end{picture}
$\
$

\hfill
\begin{minipage}{86mm}
That is, $x_1 - x_0$ is close to zero where $x_1 = \max \varphi$,
$x_0 = \min \varphi$. Therefore the periodic trajectory of (\ref{1.3})
corresponding to $\varphi$ is entirely contained in a small
neighborhood of a center $(x_c, 0)\in\mathbb{R}^2$, see Proposition
\ref{p14}. Let us recall that the period function has an expansion
\end{minipage}

\vspace{2ex}
\noindent
$$
T(\xi) = T_c + T_{2k}\varepsilon^{2k} + T_{2k+1}\varepsilon^{2k+1} +
T_{2k+2}\varepsilon^{2k+2} + \ldots
$$
with respect to $\varepsilon$, the distance between the center at
$(x_c, 0)$ and the intersection point of the orbit with the
$x$-axis $(x_0, 0)$ (see Figure 2, where the case $x_c<1$ is depicted).
The series begins always with
an even-degree coefficient $T_{2k}$ for some $k = 1, 2, \ldots$
which is called the $k$ th isochronous constant. Namely,
$T_{2k}=0$ implies also $T_{2k+1} = 0$. If all isochronous
constants vanish, all orbits around the center have the same
period $T_c$ and the center is isochronous. We will need for our
purposes however another "weighted" expansion with respect to
$\eta = \varepsilon/(1 - x_c)$ which we are going to handle below.

\noindent
\begin{prop}
\label{texp}
The explicit expression of the first isochronous constant is determined
from formulas $(\ref{T})$ and $(\ref{K})$ below.
\end{prop}

\vspace{1ex}
\noindent
{\bf Proof.}
Take a small positive $\varepsilon$ and let  $x_0 = x_c - \varepsilon$.
Then using (\ref{1.4}) and (\ref{1.5}) one obtains by direct calculations
\begin{align*}
e & = H(x_0, 0)  =|1-x_c+\varepsilon|^{b-1}[d-(x_c-\varepsilon)^2] \\
 & =
|1-x_c|^{b-1}\left(1+\frac{\varepsilon}{1-x_c}\right)^{b-1}
[d-x_c^2+2\varepsilon x_c-\varepsilon^2].
\end{align*}
Then using the identity (equivalent to \ref{root})
\begin{equation}
\label{dd}
d-x_c^2=\frac{2x_c(x_c-1)}{b-1}
\end{equation}
and denoting $\eta = \varepsilon/(1-x_c)$, we derive the formula
\begin{equation}
\label{3.2}
\begin{array}{rl}
e&\displaystyle=(d-x_c^2)|1-x_c|^{b-1}(1+\eta)^{b-1}\left[1-(b-1)\eta
+\frac{(b-1)(1-x_c)}{2x_c}\eta^2\right]\\[2mm]
&=(d-x_c^2)|1-x_c|^{b-1}e_1(\eta).
\end{array}
\end{equation}
An expansion in series with respect to $\eta$ yields immediately that
\begin{equation}
\label{e1}
e_1 (\eta) = 1 + p_2 \eta^2 + p_3 \eta^3 + p_4 \eta^4 + \ldots,
\end{equation}
$$
p_{j}=\left(
\begin{array}{c} b-1 \\ j
\end{array}\right)
-(b-1)\left(
\begin{array}{c}b-1\\ j-1
\end{array}\right)
+\frac{(b-1)(1-x_c)}{2x_c}\left(
\begin{array}{c}b-1\\ j-2
\end{array}\right),\quad j \geq 2.
$$
As $x_c$ is far from zero, we see that all coefficients $p_j$ are uniformly
bounded for $b$ fixed for all centers in all regions (I)--(IX), see Fig. 1.
Moreover,
$$
p_2 = \frac{(b-1)(1-(b+1)x_c)}{2x_c}\neq 0\;\;\mbox{\rm and}\;\;
p_j=\left(\begin{array}{c}
b-1\\j-2
\end{array}\right)
\left[p_2+{\textstyle\frac{j-2}{2j}}b(b+1)\right],\;\; j\geq 3.
$$
Similarly, from $e = H(x_0,0) = H(x_1,0)$, one can calculate the function
$\psi$ in $x_1 = x_c + \psi(\varepsilon)$. Even more conveniently, taking
$\psi (\varepsilon)=(1-x_c)\Phi(\eta)$, one obtains as above the following
equation for $\Phi$:  $e_1(\eta)=e_1(-\Phi)$.
Expanding both sides, we derive by calculations the following expansion formula,
\begin{equation}
\label{Fi}
 \Phi(\eta) = \eta + q_2 \eta^2 + q_3 \eta^3 + q_4 \eta^4 + O(\eta^5)
 \end{equation}
with
$$
q_2=\frac{p_3}{p_2},\;\;q_3=\frac{p_3^2}{p_2^2},\;\;
q_4=\frac{2p_3^3-2p_2p_3p_4+p_2^2p_5}{p_2^3}.
$$

Let us denote
$$
U (x) = e |1-x|^{1-b} + x^2 - d.
$$
Then (\ref{1.5}) becomes $Y^2 = U (X)$ and the period function is determined by
\begin{equation}
\label{1.8}
T = 2\int_{x_0}^{x_1}\frac{d X}{\sqrt{U (X)}}.
\end{equation}
We perform a change of the variables
$$
X = \frac{x_1 - x_0}{2} z + \frac{x_1 + x_0}{2}
$$
in (\ref{1.8}) and obtain
\begin{equation}
\label{smya}
T =\int_{-1}^1
\frac{(x_1 - x_0) dz}{\sqrt{U (x_c + D(z, \eta))}},
\end{equation}
where
$$
D(z, \eta) = (1 - x_c)\left(\frac{\Phi(\eta) + \eta}{2} z
+ \frac{\Phi(\eta) - \eta}{2}\right):=(1 - x_c)M(z, \eta).
$$
Next, making use of (\ref{3.2}), we get
\begin{align*}
U (x_c + D) & = e|1-x_c-D|^{1-b} + x_c^2 - d + 2x_c D + D^2 \\
& =
(d-x_c^2)\left[e_1(\eta)(1-M)^{1-b}-1-(b-1)M-\frac{(b-1)(1-x_c)}{2x_c}M^2\right].
\end{align*}
 Conditions $U (x_0) = U (x_1) = 0$ imply that $U$ vanishes for both
$ M = -\eta$ and $M = \Phi(\eta)$.
Hence, using analyticity with respect to $M$, one can rewrite $U (x_c + D)$
as
\begin{equation}
\label{uu}
\begin{array}{rl}
U&=(d-x_c^2)(M + \eta)(\Phi(\eta)-M)(A_0+A_1M+A_2M^2+\ldots)\\
&=\frac14(d-x_c^2)(\eta + \Phi(\eta))^2(1-z^2)(A_0+A_1M+A_2M^2+\ldots).
\end{array}
\end{equation}
Comparing the coefficients at the corresponding degrees $M^j$, $j=0,1,2$,
we obtain the following equations for $A_j$:
$$
\begin{array}{l}
\eta \Phi A_0 = e_1(\eta)-1,\\
\eta \Phi A_1 + (\Phi-\eta)A_0=(b-1)(e_1(\eta)-1),\\
\displaystyle \eta \Phi A_2+(\Phi-\eta)A_1-A_0 = \frac{b(b-1)}{2}e_1(\eta)-\frac{(b-1)(1-x_c)}{2x_c},\\
\eta \Phi A_j+(\Phi-\eta)A_{j-1}-A_{j-2}=(-1)^j\left(
\begin{array}{c}1-b\\j\end{array}
\right)e_1(\eta),\;\; j\geq 3. \end{array}$$
By using the expansions (\ref{e1}), (\ref{Fi}) and equality $e_1(\eta)=e_1(-\Phi)$,
we calculate
$$\begin{array}{l}
{\displaystyle A_0=p_2+\frac{p_2p_4-p_3^2}{p_2}\eta^2
+\frac{p_2p_3p_4-p_3^3}{p_2^2}\eta^3 + O(\eta^4),}\\[3mm]
{\displaystyle A_1=(b-1)p_2-p_3+\frac{(b-1)(p_2p_4-p_3^2)+p_3p_4-p_2p_5}{p_2}\eta^2
+O(\eta^3),}\\[3mm]
{\displaystyle A_2=\frac{b(b-1)}{2}p_2-(b-1)p_3+p_4+O(\eta^2),}\\[3mm]
{\displaystyle A_3=\frac{b(b^2-1)}{6}p_2-\frac{b(b-1)}{2}p_3+(b-1)p_4-p_5+O(\eta).}
\end{array}$$
On the other hand, from $M=\frac12(\Phi+\eta)z+\frac12(\Phi-\eta)$ one obtains
$$
\begin{array}{l}
M=\eta(1+\frac12q_2\eta+\frac12q_3\eta^2)z+\eta^2(\frac12q_2+\frac12q_3\eta)+O(\eta^4),\\
M^2=\eta^2(1+q_2\eta)z^2+\eta^3q_2z + O(\eta^4),\\
M^3 = \eta^3 z^3+O(\eta^4).
\end{array}
$$
Therefore, by direct calculations, we can derive the expression
$$
\begin{array}{l}
A_0 + A_1 M + A_2 M^2 + A_3 M^3 + O(M^4) \\
= p_2[1+a_1 z \eta+(b_0 + b_1 z + b_2 z^2)\eta^2 +
(c_0 + c_1 z + c_2 z^2 + c_3 z^3)\eta^3 + O(\eta^4)]
\end{array}
$$
 where
 $$
 a_1=\frac{(b-1)p_2-p_3}{p_2},\;\; b_0=\frac{(b-1)p_2p_3+2p_2p_4-3p_3^2}{2p_2^2},
 \;\;b_1={\textstyle\frac12}q_2a_1,
 $$
$$
b_2=\frac{b(b-1)p_2-2(b-1)p_3+2p_4}{2p_2},\;\; c_0=q_2b_0,\;\; c_2=q_2b_2.
$$
Next, modulo odd-degree terms with respect to $z$, one obtains
\begin{equation}
\label{alf}
[1 + a_1 z \eta + \ldots]^{-1/2} = 1 +
({\textstyle\frac38}a_1^2 z^2-{\textstyle\frac12}b_0
-{\textstyle\frac12}b_2z^2)(\eta^2+q_2\eta^3)+O(\eta^4).
\end{equation}
Finally, using (\ref{dd}) and the definition of $\Phi$
we get by direct calculations
\begin{equation}
\label{coeff}
\frac14(d-x_c^2)(\eta + \Phi(\eta))^2p_2=\frac{(b+1)x_c-1}{4(1-x_c)}(x_1-x_0)^2.
\end{equation}
Therefore, by (\ref{alf}), (\ref{coeff}) and (\ref{uu}), one obtains
(modulo odd-degree terms)
$$
\frac{x_1-x_0}{\sqrt{U (x_c+D)}} = 2\sqrt\frac{1-x_c}{(b+1)x_c-1}.
\frac{1+ (\frac38 a_1^2z^2-\frac12 b_0
-\frac12 b_2z^2)(\eta^2+q_2\eta^3)+O(\eta^4)}{\sqrt{1-z^2}}
$$
and therefore by (\ref{smya})
\begin{equation}
\label{T}
T = 2 \pi \sqrt\frac{1-x_c}{(b+1)x_c-1} \biggl[ 1+K(\eta^2+q_2\eta^3)+O(\eta^4) \biggr]
\end{equation}
with
\begin{equation}
\begin{array}{rl}
\label{K}
K&=\frac{3}{16}a_1^2-\frac12 b_0-\frac14 b_2=
{\displaystyle\frac{(b^2-4b+3)p_2^2-6(b-1)p_2p_3-12p_2p_4+15p_3^2}{16p_2^2}}\\[3mm]
&={\displaystyle\frac{b[2(b-3)(b+1)^2x_c^2-9(b-2)(b+1)x_c+12(b-1)]}{48[(b+1)x_c-1]^2}}.
\end{array}
\end{equation}
$\hfill \square $

Let us recall that our aim is to obtain sequences of $2\pi/n$-periodic
solutions $\varphi_n$ satisfying appropriate bounds in Sobolev $H^s$ norms.
For that purpose, we need the following relations
\begin{equation}
\label{cond}T=\frac{2\pi}{n},\quad |\,1-x_c| = \varepsilon^{2/s}\equiv
(x_c-x_0)^{2/s}, \mbox{where} \quad s \geq 3.
\end{equation}
We first establish the existence of solutions $\varphi$
 of (\ref{fi}) satisfying (\ref{cond}). Fix $b \neq 0, \pm 1$.
\begin{prop}
\label{exists}
Given $b\neq 0,\pm 1$ and $s \geq 3$, then there is $N_0 = N_0(b,s)$ sufficiently
large, so that for any $n \geq N_0$ there exists a periodic solution
$\varphi = \varphi_n$ of (\ref{fi}) satisfying (\ref{cond}).
\end{prop}

\vspace{2ex}
\noindent
{\bf Proof.} Clearly, the period $T$ could be small only provided that
$T_c$ is small, see  (\ref{Tc}) and (\ref{T}). Therefore,
$|1-x_c|=\varepsilon^{2/s}$ is small and such is
$|\eta|= \varepsilon^{1-\frac{2}{s}}$. To calculate $K$ in (\ref{K}) at first-order
approximation, we take $x_c = 1$ to obtain
$K=\frac{1}{48}(2b^2-11b+11)$. This implies that $T = T_c = 2\pi/n$, at first-order
approximation, which by (\ref{Tc}) yields
\begin{equation}
\label{as}
x_c = 1-\frac{b}{n^2}+o(n^{-2}).
\end{equation}
Therefore,
$$
\varepsilon=\frac{|b|^{s/2}}{n^s}+o(n^{-s}),\quad
\eta = \frac{|b|^{s/2}}{bn^{s-2}}+o(n^{2-s}).
$$
Next, by (\ref{dd}),
\begin{equation}
\label{as2}
d = 1 + \frac{2b^2}{(1-b)n^2}+o(n^{-2}), \quad \Delta=b^2\left[1-\frac{2(b+1)}{n^2}+o(n^{-2})\right].
\end{equation}
We replace this value of $d$ in conditions (i)-(iii) of
Proposition \ref{p11} (neglecting the remainder $o(n^{-2})$) to
verify that all they hold, provided that $n^2>2(1+b)$. Therefore,
Proposition \ref{p11} holds as long as $n\geq N_0=N_0(b,s)$ and
$N_0$ is sufficiently large. To verify Proposition \ref{p13} we
need to calculate $\Sigma$ in any of the cases and check that
$e\in\Sigma$. Unfortunately, first-order approximations do not
suffice to verify Proposition \ref{p13}. For that reason, we can
proceed as follows. Using the above asymptotical values, we
conclude that solutions of small period $\varphi_n$ can exist only
for parameters $b,d$ in domains I (right period annulus), II, III
and V, see the bifurcation diagram on Figure 1. Therefore, in
domains I and V, it suffices to check that
$\sqrt{d}<x_0=x_c-\varepsilon$ because $(\sqrt{d},0)$ is the
intersection point of the right branch of the invariant hyperbola
$y^2-x^2+d = 0$ with the abscissa. At first-order approximation,
this inequality is equivalent to
$$
\sqrt{d}<x_c \quad\Leftrightarrow\quad 1+\frac{b^2}{(1-b)n^2}<1-\frac{b}{n^2}
$$
which clearly holds if $|\,b|>1$. It remains to consider domains II and III
where $|\,b| < 1$, $d>1$.
The function $H(x,0)$ then has just two critical points $x_c$ and $x_s$
(a minimum at $x_c$ and a maximum at $x_s$) corresponding to the center and
the saddle. Moreover,
$$
1<x_c<x_s\quad \mbox{\rm in II},\qquad x_s<x_c<1\quad \mbox{\rm in III}.
$$
In both cases, $H (x,0)$ goes to infinity as $x \to 1$ and to minus infinity
as $|x| \to \infty$. This information implies that it suffices to prove only
\begin{equation}
\label{cnd}
e = H(x_0,0) < H(x_s,0) = e_s\quad \mbox{\rm in II, III},\quad
x_0 > 1\;\; \mbox{\rm in II},\quad x_0 > x_s \;\;\mbox{\rm in III}.
\end{equation}
By (\ref{root}), one obtains
$$
x_s = \frac{1-b}{1+b}+\frac{b}{n^2}+o(n^{-2}),
$$
by (\ref{3.2}), (\ref{e1}) we have
$$
e = \frac{2b}{(1-b)n^2}\left|\frac{b}{n^2}\right|^{b-1} \left[ 1 + o(1) \right]
$$
and by (\ref{1.4})
$$
e_s = \frac{4 b|\,2 b|^{b-1}}{(1+b)^{1+b}} \left[ 1 + O(n^{-2}) \right].
$$
As $x_0 = x_c$ at first order approximation, all conditions in (\ref{cnd})
are obviously satisfied. Thus, Proposition \ref{exists} is proved.

$\hfill \square $

So, the solutions $\varphi = \varphi_n (\tau)$ we just constructed have high frequency
since $|1 - x_c|$ is close to zero. This fact will be used in what follows.

Our next goal is to obtain simple estimates in terms of $x_c$ for the
period of the periodic solutions $\varphi$ having sufficiently small
amplitude and high frequency.
For $b \neq 0,\pm 1$ an arbitrary but fixed number, such solutions exist
in domains I, II, III and V, provided that both $d$ and $x_c$ are close
enough to 1 (as shown above). By (\ref{as}), one obtains immediately
$$\frac{|b|}{4n^2}\leq|1-x_c|\leq\frac{4|b|}{n^2},\quad n\geq N_0(b,s)$$
as long as $N_0$ is large enough. This is obviously equivalent to
\begin{equation}
\label{period1}
\pi\frac{|1-x_c|^{1/2}}{|b|^{1/2}}\leq T\leq 4\pi\frac{|1-x_c|^{1/2}}{|b|^{1/2}},
\quad n\geq N_0(b,s).
\end{equation}
Below, we write $T \simeq |1 - x_c|^{1/2}$ for the sake of (\ref{period1}).

Finally, let us rewrite equation (\ref{1.5}) in the form
\begin{equation}
\label{equ}
\varphi'^2 = \varphi^2 - d + e(1-\varphi)^{1 - b} = U (\varphi), \quad ' = d / d \tau.
\end{equation}
We shall need also the derivatives in the next section
\begin{equation}
\label{drv}
\varphi'' = \varphi+\frac{e(b-1)}{2(1-\varphi)^b}=\frac12 U'(\varphi),\qquad
\varphi''' = \left[1 + \frac{eb(b-1)}{2(1-\varphi)^{b+1}}\right]\varphi' =
\frac12 U'' (\varphi)\varphi'.
\end{equation}
Up to now we have seen that equation (\ref{equ}) admits a nonconstant even
$T$ - periodic solution (in the corresponding domains of $(b,d)$) which solves
the initial value problem
$$
\varphi'' = \varphi + \frac{e(b-1)}{2(1 - \varphi)^b}, \quad
\varphi (0) = x_0 = x_c - \varepsilon, \quad \varphi' (0) = 0.
$$

\vspace{5ex}

We conclude this section with an estimate for the
incomplete period, proceeding in the same way as above.
Take $\alpha \in \left(0, \frac{x_1 - x_0}{x_c - x_0} \right)$ and
denote
$$
\tau (x_0 + \alpha \varepsilon) =
\int_{x_0}^{x_0 + \alpha \varepsilon} \frac{d X}{\sqrt{U (X)}}.
$$
Applying the same change of variables, we obtain with
$$
\zeta = \frac{2\alpha \varepsilon}{x_1 - x_0} -1 \in (-1,1)
$$
the formula (instead of (\ref{smya}))
$$
\tau (x_0 + \alpha \varepsilon) =\frac12\int_{-1}^\zeta
\frac{(x_1-x_0)dz}{\sqrt{U(x_c + D(z, \eta))}}.
$$
Then, including in the calculation of (\ref{alf}) all terms up to
$O(\eta^2)$, we obtain
$$
\textstyle[1+a_1 z \eta + \ldots]^{-1/2}= 1 - a_1 z(\frac12\eta + \frac14q_2\eta^2)
+(\frac38 a_1^2z^2-\frac12b_0 -\frac12 b_2z^2)\eta^2+O(\eta^3),
$$
 instead. Calculating the elementary integral, we get
\begin{equation}
\label{incomplete}
\begin{array}{rl}
 \tau (x_0 + \alpha \varepsilon)&={ \displaystyle\sqrt\frac{1-x_c}{(b+1)x_c-1}}
 \biggl\{ \left(1 + G_1\eta^2\right) \left(\frac12\pi + \arcsin \zeta \right) \\[2mm]
 & + \left[ a_1(\frac12\eta+\frac14q_2\eta^2)-\zeta G_2 \eta^2 \right]
 \sqrt{1-\zeta^2} + O(\eta^3) \biggr\},
\end{array}
\end{equation}
where $G_1 = \frac{3}{16} a_1^2-\frac12b_0 -\frac14 b_2, G_2 =
\frac{3}{16} a_1^2 -\frac14  b_2$.
Recall that
$$
x_1 - x_0 = (1-x_c) (2 \eta + q_2 \eta^2 + q_3 \eta^3 + \ldots).
$$
So, we have
$$
\zeta = -1 + \frac{\alpha}{1 + \frac{q_2}{2} \eta + \frac{q_3}{2} \eta^2
+ \ldots} =
\zeta_0 + \zeta_1 \eta + O (\eta^2) ,
$$
where $ \zeta_0 = \alpha - 1, \zeta_1 = - \alpha q_2/ 2$.
Substituting this expression into (\ref{incomplete}) gives
\begin{align}
\label{incomplete1}
\tau (x_0 + \alpha \varepsilon) & = \frac{| \,1-x_c |^{1/2}}{\Delta^{1/4}}
 \Biggl\{\frac{\pi}{2} + \arcsin \zeta_0
 + \eta \left( \frac{\zeta_1}{\sqrt{1 - \zeta_0 ^2}} +
 \frac{a_1\sqrt{1 - \zeta_0 ^2}}{2} \right)  + O(\eta^2) \Biggr\} .
\end{align}
\noindent
Again, by analyticity argument we can take $\eta$ or $\varepsilon$ small enough
that the expression in the brackets in (\ref{incomplete1}) can be estimated as
follows
\begin{equation}
\label{estincomp}
\frac{|1 - x_c|^{1/2}}{2\Delta^{1/4}}\left(\frac{\pi}{2} + \arcsin (\alpha - 1)\right)
 \leq \tau (x_0 + \alpha \varepsilon) \leq
2\frac{|1 - x_c|^{1/2}}{\Delta^{1/4}}\left(\frac{\pi}{2} + \arcsin ( \alpha - 1)\right).
\end{equation}
We shall use this estimate later in section 4.

\section{Non-uniform continuity}

In this section we establish appropriate estimates in Sobolev norms of the
periodic solutions $\varphi$ derived in the previous section and then prove
our main theorem. The proof of Theorem \ref{thm1} proceeds in the line of
\cite{HM}.

Below, we will use the notation introduced in the previous sections.
In the proof of our main theorem, we are going to exploit the properties
of small-amplitude high-frequency periodic solutions $\varphi$.
\vspace{2ex}

First, let us choose the parameter $b \neq 0, \pm 1$ and freeze it.
 Next, we choose  $x_c$ so that $|\,1 - x_c |$ is sufficiently small.
And, finally, we choose a periodic orbit sufficiently close to the center
$(x_c, 0)$. That is, we choose the parameter $e$ in (\ref{1.5}) be so close to
$e_c$ in order to ensure that the amplitude $x_1 - x_0$ of the corresponding
periodic solution $\varphi$ will satisfy $x_1 - x_0 <\!<1 - x_c$.
Therefore,
\begin{equation}
  \label{4.0}
\varepsilon <\! <1 - x_c, \quad
| \varphi-x_c | \leq | x_1 - x_0 | ;\;\mbox{\rm and} \;\;
\frac{|\varphi-x_c|}{|1 - x_c|} <\!< 1.
\end{equation}


In the sequel we need $U$ and several its derivatives evaluated at $x_c$.
Trivial calculations give
\begin{align}
\label{evalu}
U (x_c) & = P \varepsilon^2 + Q \varepsilon^3 + R \varepsilon^4 + O (\varepsilon^5), \\
U' (x_c)& = \frac{b-1}{1-x_c} U (x_c), \quad U'' (x_c) = -2 P +
\frac{b (b-1)}{(1-x_c)^2} U (x_c), \quad \mbox{etc.} \ldots, \nonumber
\end{align}
where
$$\begin{array}{l}\displaystyle
P = \frac{(b+1)x_c-1}{1-x_c},\quad
 Q = \frac{(b+1)(2b-3)x_c-3(b-1)}{3(1-x_c)^2},  \\[4mm]
 \displaystyle
R= \frac{(b-2)[(b+1)(b-2)x_c-2(b-1)]}{4(1-x_c)^3}.
\end{array}$$

We begin with $L^{\infty}$ - estimates of the derivatives.
       \begin{lm}
         \label{l1}
         There exist constants $C_k (b)$, $k \in \mathbb{N}$, so that the
         following estimates hold
    \begin{equation}
    \label{221}
    |\varphi^{(k)}| \leq C_k (b) \frac{\varepsilon}{|\,1-x_c |^{k/2}}.
    \end{equation}
    \end{lm}
\noindent
{\bf Proof}.
Recall the equation $ \varphi'^2 = U (\varphi)$ and its derivatives (\ref{drv}).
Expanding $U$ around $x_c$ and using the values of $U$ and its derivatives at
$x_c$ we calculated earlier in (\ref{evalu}), we obtain
\begin{align}
|U (\varphi)|& \leq |\,U (x_c)| + |\, \varphi-x_c | |\,U'(x_c)| + \frac12|\, \varphi-x_c |^2 |\,U''(x_c)|
 + O(|\, \varphi-x_c |^3)  \nonumber \\
& \leq \left[ 1+ |\, \varphi-x_c|\frac{|\,b-1|}{|\,1-x_c |}+|\,\varphi-x_c|^2 \frac{|\, b||\,b-1|}{2 |\,1-x_c|^2}\right]
\left[ P\varepsilon^2 + |\,Q|\varepsilon^3 + O(\varepsilon^4)\right] \nonumber \\
&+\, P|\,\varphi-x_c |^2 + O(|\,\varphi - x_c |^3)   \nonumber \\
& \leq  P \varepsilon^2 \left(5 + |\,b-1| + \frac{|\,b||\,b-1|}{2} \right) + O (\varepsilon^3)
\leq  C (b) P \varepsilon^2 \nonumber
\end{align}
because of (\ref{4.0}). Since $P < |\,b|/|\,1 - x_c |$, we obtain the estimate
$$
|\, \varphi'|\leq C_1 (b)\frac{\varepsilon}{|\,1-x_c|^{1/2}}.
$$
In a similar way, developing $U' (\varphi)$, we verify the estimate
$$
|\, \varphi''|\leq C_2 (b)\frac{\varepsilon}{|\,1-x_c |}.
$$
Next, we are going to proceed by induction.
Taking $k$th-order derivative of the both sides of (\ref{fi}),
$k = 0, 1, 2, \ldots$, we obtain the equation
\begin{equation}
\label{kth1}
(1 - \varphi) \varphi^{(k+3)} = \varphi^{(k+1)} +
\sum\limits_{i=0}^k \left[ c_i\varphi^{(i+1)}\varphi^{(k-i+2)} + d_i\varphi^{(i)}
\varphi^{(k-i+1)}\right],
\end {equation}
where $c_i, d_i$ are certain constants depending on $k$ and $b$.
Applying the induction hypothesis and the first bound from (\ref{4.0}),
we conclude that
$$
|\,1 - \varphi||\varphi^{(k+3)}|\leq
C_k (b)\frac{\varepsilon}{|\,1-x_c |^{(k+1)/2}}.
$$
As $|\,1 - \varphi | > |\, 1 - x_c - O(\varepsilon) | > \frac12 |\,1 - x_c|$,
the claim follows.

$\hfill \square $

Next we turn to $L^2$ - estimates.

       \begin{lm}
         \label{l2}
There exist constants $D_k (b)$, $k \in \mathbb{N}$, so that the following
estimates hold
\begin{equation}
 \label{2221}
 ||\,\varphi'||_{L^2[-\frac{T}{2}, \frac{T}{2}]} ^2
              \leq D_1 (b) \frac{\varepsilon^2}{|\,1-x_c |^{1/2}}.
\end{equation}
and for any $k = 2, 3, \ldots$
 $$
 ||\,\varphi^{(k)}||_{L^2[-\frac{T}{2}, \frac{T}{2}]} ^2 \leq
 D_k (b) \frac{||\,\varphi'||^2 _{L^2[-\frac{T}{2},\frac{T}{2}]}}{|\,1-x_c |^{(k-1)}}.
   $$
    \end{lm}
\noindent
{\bf Proof}. For the first derivative, we have
$$
||\, \varphi' ||^2 _{L^2[-\frac{T}{2}, \frac{T}{2}]} =
\int_{-\frac{T}{2}} ^{\frac{T}{2}} \varphi'^2 d \tau \leq
C_1 ^2 (b) \frac{\varepsilon^2}{|\,1-x_c |} T
\leq D_1 (b) \frac{\varepsilon^2}{|\,1-x_c |^{1/2}}.
$$
Next, we get by (\ref{drv}), (\ref{3.2})  and
$| 1 - \varphi | > \frac12| 1 - x_c |$
$$
||\varphi''||^2_{L^2[-\frac{T}{2}, \frac{T}{2}]} =
\int_{-\frac{T}{2}} ^{\frac{T}{2}}  \varphi''^2 d \tau =
-\int_{-\frac{T}{2}} ^{\frac{T}{2}} \varphi'''\varphi' d \tau =
-\frac12 \int_{-\frac{T}{2}} ^{\frac{T}{2}} U''(\varphi)\varphi'^2 d \tau =
$$
$$
 \int_{-\frac{T}{2}} ^{\frac{T}{2}} \left(
\frac{b(1-x_c)^b}{(1-\varphi)^{b+1}} e_1 (\eta)  - 1 \right) \varphi'^2 d \tau \leq
   D_2 (b) \frac{||\varphi'||^2 _{L^2}}{|\, 1-x_c |}.
$$
Finally, we again proceed by induction. Lemma \ref{l2} holds for $k = 2$.
By using (\ref{kth1}), (\ref{221}) and the inductive hypothesis, one easily
obtains
\begin{align}
 &||\,(1-\varphi)\varphi^{(k+3)}||\leq ||\, \varphi^{(k+1)}||+
\sum\limits_{i=0}^k(c_i||\varphi^{(i+1)}\varphi^{(k-i+2)}||+d_i||\, \varphi^{(i)}
\varphi^{(k-i+1)}||) \leq \nonumber \\
 & \left[
\frac{D_{k+1}}{|\,1-x_c|^{k/2}}+
\sum\limits_{i=0}^k\left(\frac{c_iD_{i+1}}{|1-x_c|^{i/2}}.\frac{C_{k-i+2}\varepsilon}
{|1-x_c|^{(k-i+2)/2}}+ \frac{d_iC_i\varepsilon}{|1-x_c|^{i/2}}.\frac{D_{k-i+1}}
     {|1-x_c|^{(k-i)/2}}\right)\right]||\, \varphi'|| \nonumber \\
&\leq \frac{D_{k+3}}{|1-x_c|^{k/2}} ||\,\varphi'||  \nonumber.
\end{align}
As  $||\,(1 - \varphi)\varphi^{(k+3)}||\geq\frac12|1-x_c| ||\,\varphi^{(k+3)}||$,
the statement follows by induction.

$\hfill \square $

Recall the Sobolev norm

$$
||\, f ||^2 _{\mathrm{H}^s} =
\sum_{\xi \in \mathbb{Z}} (1 + \xi^2)^s |\, \hat{f} (\xi) |^2 ,
$$
where $\hat{f} (\xi)$ is the Fourier transform of $f$.

  \begin{lm}
    \label{l3}
      Let $\varphi = \varphi_n$ be the $T = \frac{2\pi}{n}$-periodic solution
      constructed in the end of the previous section.
      For any $s \geq 3$, there is a
      positive constant $c_{s ,b}$ depending only on $s$ and $b$, such that
 $$
 ||\, \varphi ||^{2}_{\mathrm{H}^{s}{(-\pi, \pi)}} \leq
   c_{s, b}\left( {\frac{1}{| 1 - x_c |^{s-1}}}||\, \varphi'||^{2}_{L^{2}(-\pi, \pi)}
   + x_1^{2}\right).
   $$
  \end{lm}
\noindent
    {\bf Proof.}
    Let $s = k$, where $k = 3, 4,...$. Using the facts that
      $$
      ||\, \varphi^{(k)}||^{2}_{L^{2}(-\pi, \pi)} =
      n||\, \varphi^{(k)}||^{2}_{L^{2}({\frac{-\pi}{n}}, {\frac{\pi}{n}})} \quad
      \mbox{and} \quad  \, x_0 \leq \varphi \leq x_1 $$
      these estimates follow from Lemma \ref{l2}.

Let now $s = k + \sigma$, where $k \geq 3$ is a positive integer and
$0 < \sigma < 1$. We follow Proposition 3.3 in \cite{HM}
$$
||\, \varphi ||^2 _{\mathrm{H}^{s}{(-\pi, \pi)}} \lesssim
||\, \varphi^{(k)} ||^2 _{\mathrm{H}^{\sigma}{(-\pi, \pi)}} +
||\, \varphi' ||^2 _{L^2 {(-\pi, \pi)}} + ||\, \varphi||^2 _{L^2 {(-\pi, \pi)}}.
$$
 We have $||\, \varphi||^2 _{L^2 {(-\pi, \pi)}} =
 n ||\, \varphi||^2 _{L^2 {(-\pi/n, \pi/n)}} \backsimeq 2 \pi x_1 ^2$. It
 remains to estimate the $\mathrm{H}^{\sigma}$ - norm of $\varphi^{(k)}$.
 It is proven in
 \cite{HM} that for any smooth $f$ the following inequality holds
     $$
 ||\,f||_{\mathrm{H}^{\sigma}(-\pi, \pi)} \lesssim
||\,f||^{1-\sigma}_{L^{2}(-\pi, \pi)} ||\,f||^{\sigma}_{\mathrm{H}^{1}(-\pi, \pi)}.
     $$
Applying this to $\varphi^{(k)}$ yields
$$
 ||\, \varphi^{(k)} ||_{\mathrm{H}^{\sigma}(-\pi, \pi)} \lesssim
 ||\, \varphi^{(k)} ||^{1-\sigma}_{L^{2}(-\pi, \pi)}
 ||\, \varphi^{(k)} ||^{\sigma}_{\mathrm{H}^{1}(-\pi, \pi)}.
$$
Since $| 1- x_c| < 1$, using the estimates from Lemma \ref{l2} we obtain
     $$
   ||\, \varphi^{(k)}||_{\mathrm{H}^{1} (-\pi, \pi)} \backsimeq
   ||\, \varphi^{(k)}||_{L^2 (-\pi, \pi)} + ||\, \varphi^{(k+1)}||_{L^2 (-\pi, \pi)} \lesssim
     $$
   $$
   \left(\frac{1}{|1-x_c|^{(k-1)/2}} + \frac{1}{|1-x_c|^{k/2}} \right)
   ||\,\varphi'||_{L^{2} (-\pi, \pi)} \lesssim
   \frac{||\,\varphi'||_{L^{2} (-\pi, \pi)}}{|1-x_c|^{k/2}}.
   $$
   Combining these inequalities, we get
   $$
      ||\,\varphi^{(k)}||_{\mathrm{H}^{\sigma}(-\pi, \pi)}    \lesssim
\frac{||\,\varphi'||^{1-\sigma} _{L^{2} (-\pi, \pi)}}{|1-x_c|^{(k-1)(1-\sigma)/2}}
\frac{||\,\varphi'||^{\sigma} _{L^{2} (-\pi, \pi)}}{|1-x_c|^{k \sigma/2}}
  \lesssim
\frac{||\,\varphi'||_{L^{2} (-\pi, \pi)}}{|1-x_c|^{(k + \sigma-1)/2}}
  $$
    from where the lemma follows.

    $\hfill \square $

\vspace{2ex}

\noindent
{\bf Proof of Theorem \ref{thm1}.}
Let $s \geq 3$ and let $\varphi_n$ be the $2 \pi/n$ - periodic smooth
solution, constructed above. Recall from section 3 that
\begin{equation}
\label{p1}
n \backsimeq \frac{1}{|1 - x_c|^{1/2}}.
\end{equation}
Consider the following two sequences of travelling wave solutions
\begin{equation}
\label{p2}
  u_n (x, t) = \varphi_n (x - t), \qquad v_n (x, t) = c_n \varphi_n
(x - c_n t)
\end{equation}
and take
$$
c_n = 1 + \frac{1}{n}.
$$
As in \cite{HM} we  show that these sequences are bounded, their
difference goes to zero at time $t = 0$ and stays apart from zero at $t > 0$.

The boundedness  and the limit at the time $t = 0$ are almost
straightforward. Taking into account
(\ref{2221}), (\ref{p1}) it is obtained
$$
||\, \varphi'_n||^{2}_{L^{2}(-\pi, \pi)} =
      n||\, \varphi'_n||^{2}_{L^{2}({\frac{-\pi}{n}}, {\frac{\pi}{n}})}
      \lesssim \frac{1}{|1-x_c|^{1/2}} \frac{D_1 (b)\varepsilon^2}{|1-x_c|^{1/2}}.
      $$
Also, we have from (\ref{p2}) and Lemma \ref{l3} that
$$
|| v_n (t) ||^{2}_{\mathrm{H}^s(-\pi, \pi)}=
c^2_n ||\, \varphi_n||^{2}_{\mathrm{H}^s(-\pi, \pi)}
\lesssim c_n^{2}c_{b,s} \frac{\varepsilon^2}{|1-x_c|^s} + x_1 ^2,
$$
where $s \geq 3$. The choice of parameters $|1-x_c|^s =
\varepsilon^2$ assures that the both sequences $u_n$ and $v_n$ are bounded
in $\mathrm{H}^s$ - norms.

Further,
 $$
 ||\, v_n (0) - u_n (0) ||^2 _{\mathrm{H}^s ( \mathbb{S})} =
 ||\, c_n \varphi_n - \varphi_n ||^2 _{\mathrm{H}^s (\mathbb{S})} =
 (c_n - 1)^2||\, \varphi_n ||^2 _{\mathrm{H}^s (\mathbb{S})} \cong \frac{1}{n^2},
 $$
which goes to $0$ when $n \to \infty$.

Finally, the behavior at time $t > 0$ can be established in the following way.
$$
 ||\, v_n (t) - u_n (t) ||^2 _{\mathrm{H}^s (\mathbb{S})}=
 \sum_{\xi \in \mathbb{Z}}(1+\xi^2)^s|\,c_n\widehat{\varphi_n}(\cdot-c_nt)(\xi) -
 \widehat{\varphi_n}(\cdot-t)(\xi)|^2,
$$
where $\widehat{\varphi_n}(\cdot-c_nt)(\xi)$ is the Fourier transform of the
function $ \varphi_n (x - c_nt)$ with respect to $x$, that is
after changing the variables
$\widehat{\varphi_n}(\cdot-c_n t)(\xi)=e^{-itc_n\xi}\widehat{\varphi_n}(\xi)$.
Hence,
$$
 ||\, v_n (t) - u_n (t) ||^2 _{\mathrm{H}^s (\mathbb{S})}=
 \sum_{\xi\in \mathbb{Z}}(1+\xi^2)^s\left| (e^{\frac{-it\xi}{n}}-1)
 + {\frac{1}{n}}e^{\frac{-it\xi}{n}}\right|^2 |\,\widehat{\varphi}(\xi)|^2,
$$
and
  $$
  ||\, v_n (t) - u_n (t) ||^2 _{\mathrm{H}^s (\mathbb{S})} \geq
     (1 + n^2)^s | (e^{-i t} - 1) + \frac{1}{n} e^{-i t}
     |^2 |\, \hat{\varphi}_n (n)|^2 .
     $$
Since $\varphi_n$ is a $2\pi/n$ - periodic, even function and after
integrating by parts, we get
$$
\hat\varphi_n (n) = \frac{n}{\sqrt{2 \pi}} \int_{-\pi/n} ^{\pi/n}
e^{-i n \tau} \varphi_n (\tau) d \tau =
-\frac{2}{\sqrt{2 \pi}} \int_0 ^{\pi/n} \sin (n \tau) \varphi' (\tau) d \tau.
$$
Therefore
\begin{equation}
\label{p5}
||\, v_n (t) - u_n (t) ||^2 _{\mathrm{H}^s (\mathbb{S})} \geq \frac{2}{\pi}
     (1 + n^2)^s | (e^{-i t} - 1) + \frac{1}{n} e^{-i t}
     |^2 |\, B_n |^2 ,
\end{equation}
where we denote
$$
 B_n = \int_0 ^{\pi/n} \sin(n \tau) \varphi'_n (\tau) d \tau .
$$


\noindent
 The integral for $B_n$ can be estimated from below in the same line as
 Lemma 4.1 in \cite{HM}.

\begin{lm}
\label{l4}
There exists a constant $c_0  > 0$ independent of n such that
$$
  B_n \geq c_0 \varepsilon.
$$
\end{lm}
\noindent
{\bf Proof.} We have
$$
 B_n = \int_0 ^{\pi/n} \sin(n \tau) \varphi'_n (\tau) d \tau  =
 \int_{x_0} ^{x_1} \sin(n \tau ( \varphi)) d \varphi .
$$
For any  $\alpha \in (0, \frac{x_1 - x_0}{x_c - x_0} )$,
\begin{equation}
\label{p6}
B_n \geq
\int_{x_0 + \alpha \varepsilon/2} ^{x_0 + \alpha \varepsilon} \sin(n \tau ( \varphi)) d \varphi .
\end{equation}
We take $\alpha$ to satisfy the condition
\begin{equation}
\label{p7}
n \tau ( x_0 + \alpha \varepsilon) \leq \frac{\pi}{2}.
\end{equation}
To do this, let us first recall the estimate (\ref{period1}) on
the period. Next, by (\ref{as2}), we have
$\Delta = b^2[1+O(n^{-2})]$, therefore one can rewrite (\ref{period1}) as
$$
\pi \frac{|1 - x_c|^{1/2}}{\Delta^{1/4}} \leq T  = \frac{2 \pi}{n} \leq
4 \pi \frac{|1 - x_c|^{1/2}}{\Delta^{1/4}}.
$$
Further, taking advantage from the estimate on the incomplete period
(\ref{estincomp}), we get
\begin{equation}
\label{p8}
 \frac{1}{4}\left(\frac{\pi}{2} + \arcsin ( \alpha - 1)\right)  \leq
        n \tau ( x_0 + \alpha \varepsilon) \leq
        4 \left(\frac{\pi}{2} + \arcsin ( \alpha - 1)\right).
\end{equation}
Thus, to satisfy the condition (\ref{p7}) we take $\alpha$ so that
$$
4 \left(\frac{\pi}{2} + \arcsin ( \alpha - 1)\right) = \frac{\pi}{2},
$$
or $\arcsin (\alpha-1) = -\frac{3 \pi}{8}$. With this choice of $\alpha$
inequality (\ref{p6}) gives
\begin{align}
B_n & \geq
\int_{x_0 + \alpha \varepsilon/2} ^{x_0 + \alpha \varepsilon}
\sin(n \tau (x_0 + \frac{\alpha}{2} \varepsilon )) d \varphi  \nonumber \\
& = \sin \left(n \tau (x_0 + \frac{\alpha}{2} \varepsilon ) \right)
\frac{\alpha}{2} \varepsilon \geq \left[ \frac{\alpha}{2}
\sin \left( \frac{1}{4}\left(\frac{\pi}{2} +
\arcsin ( \frac{\alpha}{2} - 1)\right) \right) \right] \varepsilon, \nonumber
\end{align}
where the last inequality follows from the lower bound in (\ref{p8}) and $\alpha$
is replaced by $\alpha/2$. This proves the lemma.

$\hfill \square $

Returning to (\ref{p5}) one gets
$$
  ||\, v_n (t) - u_n (t) ||^2 _{\mathrm{H}^s (S)} \gtrsim
    n^{2 s} \varepsilon^2 \left | (e^{-i t} - 1) + \frac{1}{n} e^{-i t} \right |^2.
$$
Thus, the desired estimate is obtained as in \cite{HM} using
(\ref{p1}) and $|1 - x_c|^s = \varepsilon^2$.

$\hfill \blacksquare $

\section{The cases $\mathrm{b} = \pm 1$.}

Here we study the cases $b = \pm 1$ in the Holm - Staley equation (\ref{0.1}).
Since most of the computations and estimates are similar to those in sections
2, 3 and 4, we give only the key results and differences.

\subsection{The case $\bf b = -1$.}

Equation (\ref{0.4}) with $b=-1$ has no
hydrodynamical relevance, but we consider it here due to its simplicity.
One should start with it, because  all things are transparent.
By (\ref{1.2}), we obtain the conic curve
\begin{equation}
\label{b.33}
(\varphi-1)^{-2}(\varphi'^2 -\varphi^2+d)=e.
\end{equation}
There are periodic solutions $\varphi$ for $d>1$ and
$e \in \left(\frac{d}{1-d},-1\right)$.
They surround the center at $(d, 0)$. One can rewrite (\ref{b.33}) as
$$
\varphi'^2 + \frac{1}{e}(\varphi-1)^2 - 2(\varphi-1) + d =0,
$$
with new $d$ and $e$ (equal to $d-1$ and $-(e+1)^{-1}$, respectively).
Hence, periodic solutions exist for  $e > d > 0$. Denote
$A = \sqrt{e(e-d)}$. Then they are given explicitly by
  $$
  \varphi(\tau) = 1 + e - A \cos \frac{\tau}{\sqrt{e}}
  $$
with period $T = 2 \pi \sqrt{e}$.
Assuming $A$ and $e$ small, one can find integer $n$ such that
$n \simeq \frac{1}{\sqrt{e}}$ and $ T = \frac{2 \pi}{n}$.

As before, let us take the following two sequences of solutions
$$
u_n (x, t) = \varphi_n (x - t), \quad v_n (x, t) = c_n \varphi_n (x - c_n t),
\quad c_n = 1 + \frac{1}{n}.
$$
It is sufficient to estimate $v_n$. A direct computation gives
$$
||\,v_n (t) ||_{\mathrm{H}^s(-\pi, \pi)}^2= c_n^2 \left[
(1+e)^2 + {\frac{1}{4}}(1+n^2)^s A^2 \right] \leq
4 \left [(1+e)^2 + 2^{s-2} n^{2 s} A^2 \right].
$$
Boundedness is achieved upon the condition $A^2 = e^s, s \geq 3$.
The limit at $t = 0$ is the same as above.
It remains to consider the estimate
$$
||\, v_n (t) - u_n (t) ||^2 _{\mathrm{H}^s (\mathbb{S})} \geq
     (1 + n^2)^s | (e^{-i t} - 1) + \frac{1}{n} e^{-i t}
     |^2 |\, \hat{\varphi}_n (n)|^2 .
$$
Trivial computations yield that $\displaystyle \hat{\varphi}_n (n) =
- A/2$,
 so
$$
||\, v_n (t) - u_n (t) ||^2 _{\mathrm{H}^s (\mathbb{S})} \geq
     \frac{1}{4} A^2 n^{2 s} | (e^{-i t} - 1) + \frac{1}{n} e^{-i t} |^2
     = \frac{1}{4} | (e^{-i t} - 1) + \frac{1}{n} e^{-i t} |^2
$$
in view of relation $A^2 = e^s, s \geq 3$. Hence, the result follows as in
\cite{HM}.

\subsection{Case $\bf b = 1$.}

By (\ref{b=1}),
travelling-wave solutions of (\ref{0.1}) of the form
$u = y (x - t)$, $y < 1$ will satisfy
\begin{equation}
\label{a1}
H(y,y') \equiv y'^2 - y^2 - 2 d \ln (1 - y) = e.
\end{equation}
Hence, we have a conservative system with a Newtonian first integral $H$.
The critical points of the potential $H (y, 0)$ are
$$
y_c = \frac{1+\sqrt{1 - 4 d}}{2}, \quad y_s = \frac{1-\sqrt{1 - 4 d}}{2},
\quad d < 1/4,
$$
where $c$ stands for the center and $s$ for the saddle.
It is straightforward to verify that for $d \in (0, 1/4)$ and
$H (y_c,0) < e < H (y_s,0)$ there
are periodic solutions (see also Figure 3).
\begin{figure}
\label{fig3}
\addtocounter{figure}{1}
\begin{center}
 \setlength{\unitlength}{1.9mm}
 \begin{picture}(35,45)(0,0)
 \linethickness{0.4pt}

       \qbezier(0,34)(15,34)(23,34)  
       \qbezier(0,31)(15,31)(35,31)  

       \put(-2,34){\makebox(0,0)[cc]{$e$}}
       \put(35,32.5){\makebox(0,0)[cc]{$y$}}
       \qbezier(35,31)(34,30.5)(33,30)   
       \qbezier(35,31)(34,31.5)(33,32)   
       \qbezier(0,30)(0,38)(0,43)    

       \put(19,10){\makebox(0,0)[cc]{${\scriptstyle 1}$}}
       \qbezier(18,10)(18,13)(18,15)
       \qbezier(18,31)(18,38)(18,43)
       \put(19,30){\makebox(0,0)[cc]{${\scriptstyle 1}$}}

       \qbezier(0,43)(0.5,42)(1,41)  
       \qbezier(0,43)(-0.5,42)(-1,41)    
       \put(2,43){\makebox(0,0)[cc]{$H(y,0)$}}

       \qbezier(0,12)(15,12)(35,12)  
       \put(35,15){\makebox(0,0)[cc]{$y$}}
       \qbezier(35,12)(34,12.5)(33,13)   
       \qbezier(35,12)(34,11.5)(33,11)   
       \qbezier(0,5)(0,12)(0,19)    
       \put(2.5,20){\makebox(0,0)[cc]{$y'$}}
       \qbezier(1,17)(0.5,18)(0,19)  
       \qbezier(-1,17)(-0.5,18)(0,19)    

 \linethickness{0.8pt}
       \qbezier(5,36)(9,29)(13,34)
       \qbezier(13,34)(16,37)(17,43)
       \qbezier(0,31)(3,38)(5,36)

 \linethickness{0.8pt}

    \qbezier(6.4,12)(6.4,14)(9.3,14)
    \qbezier(9.3,14)(12.5,14)(12.7,12)
    \qbezier(9.3,10)(12.5,10)(12.7,12)
    \qbezier(10,10)(10.4,10.4)(10.8,10.8)
    \qbezier(10,10)(10.5,9.8)(11,9.6)
    \qbezier(6.4,12)(6.4,10)(9.3,10)

\put(9.9,31){\makebox(0,0)[cc] {${\cdot}$}}
\put(10,30){\makebox(0,0)[cc] {${\scriptstyle y_c}$}}

\put(8,13){\makebox(0,0)[cc] {${\scriptstyle \varepsilon}$}}

\put(9.9,12){\makebox(0,0)[cc] {${\cdot}$}}
\put(10,11){\makebox(0,0)[cc] {${\scriptstyle y_c}$}}

\put(5,10.5){\makebox(0,0)[cc] {${\scriptstyle y_0}$}}
\put(15,10.5){\makebox(0,0)[cc] {${\scriptstyle y_1}$}}
 \end{picture}
$\
$
\end{center}
\caption{\hspace{-0.5pt} The periodic solution of (\ref{a1}) }
\end{figure}

Let $y_1 = \max y$ and $y_0 = \min y$. We assume that $y_1 - y_0$ is small.
 So, there is a periodic solution to (\ref{a1}) which satisfies the
 following initial value problem
$$
y'' = y  - \frac{d}{1-y}, \quad y (0) = y_0, \quad y' (0) = 0.
$$
The period function has an expansion
$$
 T (\varepsilon) = T_c + \varepsilon^2 T_2 + \ldots,
$$
where $\varepsilon$ is defined as
$$
y_0 = y_c - \varepsilon \quad \mbox{and} \quad T_c = 2 \pi\sqrt{\frac{1-y_c}{ 2 y_c-1}}.
$$
Note that, when $d < 1/4$, then $y_c > 1/2$. Similar computations as in
Section 2 give
$$
e = H (y_0, 0) =  - y_c ^2 - 2 d \ln (1 - y_c) + P \varepsilon^2 + Q \varepsilon^3 +
R \varepsilon^4 + \ldots,
$$
where
$$
 P = \frac{2 y_c - 1}{1 - y_c}, \quad Q = -\frac{2 y_c}{3(1 - y_c) ^2},
 \quad R = \frac{y_c}{ 2 (1 - y_c) ^3}.
$$
In terms of $\eta$ the energy $e$ becomes
$$
e = H (y_0, 0) =  - y_c ^2 - 2 d \ln (1 - y_c) + p_2 \eta^2 + p_3 \eta^3 +
p_4 \eta^4 + \ldots,
$$
where
$$
 p_2 = (2 y_c - 1) (1 - y_c), \quad p_3 = -\frac{2}{3} y_c (1-y_c),
 \quad p_4 = \frac{y_c}{2} (1-y_c).
$$
For these calculations we have used the identity $y_c -  d/(1 - y_c) = 0$.
From $H (y_0, 0) = H (y_1, 0)$ one obtains that
$$
 y_1 = y_c +  (1-y_c) \Phi (\eta) , \quad \Phi (\eta) = \eta + r \eta^2 + r^2 \eta^3 + \ldots
$$
with $r = p_3/p_2$. Denote $U (y) = y^2 + 2 d \ln (1 - y) + e$. Then
(\ref{a1}) becomes $y'^2 = U (y)$ and the period function is
$$
 T = 2 \int_{y_0} ^{y_1} \frac{d y}{\sqrt{U (y)}}.
$$
As before we put
$$
y = \frac{y_1 - y_0}{2} z + \frac{y_1 + y_0}{2},
$$
thus
$$
T = \int_{-1}^1
\frac{(y_1 - y_0) \, d z}{\sqrt{U (y_c + D(z, \eta))}}.
$$
In the same line of computations we obtain the formula
\begin{equation}
\label{a51}
T = 2 \pi \sqrt{\frac{1-y_c}{2 y_c - 1}} \left( 1 +
\frac{y_c (9 - 8 y_c)}{24 (2 y_c - 1)^2}  \eta^2 + \ldots \right).
\end{equation}

As above one can take $\eta$ so small, that the expression in the brackets
in (\ref{a51}) will take values in $[\frac{1}{2}, 2]$. This gives
 $$
 \pi \sqrt{\frac{1-y_c}{2 y_c - 1}}   \leq T \leq 4 \pi  \sqrt{\frac{1-y_c}{2 y_c - 1}} .
$$
We write $T \simeq \sqrt{1-y_c}$ and for any sufficiently large integer $n$
one can find $y_c$ so that and $1 - y_c$ is sufficiently small in order to
achieve
$$
T = \frac{2 \pi}{n} \quad \mbox{and} \quad
n \simeq \frac{1}{\sqrt{1 - y_c}}.
$$
Hence, we have constructed high-frequency solution $y = y_n (t)$ with period
$T = 2 \pi / n$.

Next, in order to estimate the incomplete integral $\tau (y_0 + \alpha \varepsilon)$,
by long but straightforward computations similar to those in section 3 we obtain
$$
\frac{1}{2} \sqrt{\frac{1-y_c}{2 y_c - 1}}\left(\frac{\pi}{2} + \arcsin ( \alpha - 1) \right)
 \leq \tau (y_0 + \alpha \varepsilon) \leq
2 \sqrt{\frac{1-y_c}{2 y_c - 1}} \left(\frac{\pi}{2} + \arcsin ( \alpha - 1)\right).
$$

Finally, we need some estimates in order to obtain upper bounds for these
solutions. We need writing (\ref{a1}) in the form $y'^2 = U (y)$ and
then calculate the derivatives
\begin{equation}
\label{a6}
y'' = \frac12 U' (y), \qquad y''' = \frac12 U'' (y) y'.
\end{equation}
Assume that $y_c$ is close enough to 1 and $y_1 - y_0 << 1 - y_c$, and also
\begin{equation}
  \label{a7}
\varepsilon <\! <1 - y_c, \quad
| y - y_c | \leq | y_1 - y_0 | \;\;\mbox{\rm and} \;\;
\frac{|y - y_c|}{1 - y_c} <\!< 1.
\end{equation}
Expanding $U$ around $y_c$, using (\ref{a7}) and that $P\leq 1/(1-y_c)$ we
obtain the estimate
$$
| y' | \leq \frac{\sqrt{10} \, \varepsilon}{(1-y_c)^{1/2}}.
$$
In a similar way, developing $U' (y)$ we get
$
| y'' | \leq 4 \varepsilon /(1-y_c).
$
Again, induction arguments give the estimates
$$
| y^{(k)} | \leq C_k \frac{\varepsilon}{(1 - y_c)^{k/2}}.
$$
From the above expressions we obtain $L^2$-estimate for the
first derivative
$$
|| y' ||^2 = \int_{-T/2} ^{T/2} y'^2 d \tau
\leq C_1 \frac{\varepsilon^2}{1-y_c} T \leq D_1 \frac{\varepsilon^2}{(1-y_c)^{1/2}}.
$$
Finally, from (\ref{a6}) we obtain an estimate for the second derivative
$$
|| y''||^2_{L^2} =
\int_{-\frac{T}{2}} ^{\frac{T}{2}}  y''^2 d \tau =
-\int_{-\frac{T}{2}} ^{\frac{T}{2}} y''' y' d \tau =
-\frac12 \int_{-\frac{T}{2}} ^{\frac{T}{2}} U''(y) y'^2 d \tau
\leq D_2 \frac{|| y'||^2_{L^2}}{1-y_c}.
$$
Now, we can proceed by induction to obtain similar estimates for the
higher-order derivatives as in Lemma \ref{l2}. The proof of Theorem
\ref{thm1} is then finished in the same way as in the general case.

\section{Conclusions}

In this paper we study the Cauchy problem for the periodic
Holm - Staley b - family of equations. The results by
Himonas and Misiolek \cite{HM} (proved for the CH equation
$b = 2$ only) and the one for the DP equation $b = 3$ \cite{ChHa}, are
extended for the general case of $b$-family $b \neq 0$ (Theorem \ref{thm1}).
 We show that the solution map is not uniformly continuous in
 $\mathrm{H}^s, s \geq 3$. The proof is based on the construction of
 suitable smooth periodic solutions of small amplitude. To our knowledge,
 this idea comes from Kato \cite{Kato}.

Our result for the whole b-family is weaker than the results for particular
values of $b$ in the above mentioned papers \cite{HM} and \cite{ChHa}
where $s \geq 2$. This is because we assume that the small parameters
$\varepsilon$ and $|1 - x_c |$ are related as $\varepsilon << |1 - x_c |$.
We need this assumption in order to estimate the period, which is the
main difficulty here. Then the relation $|1 - x_c |^s = \varepsilon^2$ is
valid for $s > 2$.

From the other hand, an interpolation argument is used to obtain the estimates
for non-integer Sobolev indexes. That is the reason why the range $2 < s < 3$
is not covered. Perhaps, one should consider the case $s = 2$ separately, but
this makes the estimates of the period for arbitrary $b$ more difficult.

\end{document}